\documentclass{Interspeech}
\usepackage{amsmath}
\usepackage{cite}
\usepackage{multirow}

\interspeechcameraready

\title{Uni-VERSA: Versatile Speech Assessment with a Unified Network}

\author[affiliation={1}]{Jiatong}{Shi}
\author[affiliation={1}]{Hye-Jin}{Shim}
\author[affiliation={1}]{Shinji}{Watanabe}

\affiliation{Language Technologies Institute}{Carnegie Mellon University}{U.S.A.}
\email{jiatongs@cs.cmu.edu}

\usepackage{comment}

\begin{document}

\maketitle

\begin{abstract}
Subjective listening tests remain the golden standard for speech quality assessment, but are costly, variable, and difficult to scale. In contrast, existing objective metrics, such as PESQ, F0 correlation, and DNSMOS, typically capture only specific aspects of speech quality. To address these limitations, we introduce Uni-VERSA, a unified network that simultaneously predicts various objective metrics, encompassing naturalness, intelligibility, speaker characteristics, prosody, and noise, for a comprehensive evaluation of speech signals. We formalize its framework, evaluation protocol, and applications in speech enhancement, synthesis, and quality control. A benchmark based on the URGENT24 challenge, along with a baseline leveraging self-supervised representations, demonstrates that Uni-VERSA provides a viable alternative to single-aspect evaluation methods. Moreover, it aligns closely with human perception, making it a promising approach for future speech quality assessment.
\end{abstract}

\section{Introduction}
\label{sec:intro}


Speech profiling or speech quality assessment becomes a necessary function to evaluate various tasks, including speech synthesis, enhancement, separation, and coding~\cite{huang22f_interspeech, yi22b_interspeech, shi2024espnet, torcoli2021objective}. The task is now getting more attention due to the growing demand for reliable and scalable evaluation methods in real-world applications, driven by rapid advancements in generative speech AI models. However, the golden standard for assessing speech quality remains the listening test, which collects human opinions through methods like mean opinion score (MOS) and A/B testing.


Evaluating speech generation models through human preferences poses persistent challenges~\cite{huang22f_interspeech, cooper2024review}. A major issue is the variability among listeners, shaped by perceptual sensitivity, cultural background, and individual biases~\cite{zielinski2008some}. This variability necessitates large-scale data collection for statistical significance, making the process costly and hard to scale. Recruiting and compensating participants while maintaining response quality demands substantial resources. Crowdsourced evaluations add noise due to inattentive or malicious raters~\cite{loizou2011speech, jimenez2021removing}. Even with clear guidelines, listeners may prioritize different aspects, such as naturalness or intelligibility, when rating~\cite{naderi2020towards}. These challenges underscore the complexity of human evaluation and the importance of supplementing it with objective metrics.

Objective evaluation of speech signals is essential to overcome the limitations of human listening tests~\cite{cooper2024review, torcoli2021objective, shi2024versa}. 
Early approaches relied on signal processing principles and psychoacoustic findings to design assessment metrics such as mel cepstral distortion (MCD), PESQ, and STOI~\cite{kubichek1993mel, rix2001perceptual, taal2010short}. These metrics typically require reference signals for computation. To eliminate this dependency, research has shifted towards non-intrusive metrics that leverage neural networks to directly predict human preference. Examples include DNSMOS for noisy speech quality estimation~\cite{reddy2021dnsmos} and models like UTMOS and MOSNet for evaluating speech synthesis quality~\cite{saeki22c_interspeech}. 
In general, previous work used a single metric for speech quality~\cite{cooper2024review}, but the quality of speech is naturally multidimensional, covering intelligibility, naturalness, and background noise~, among others~\cite{koster2018multidimensional, naderi2024multi}. Therefore, single-aspect metrics risk fragmented, biased evaluations.
One approach to addressing this challenge is leveraging pre-defined metrics in each domain to provide a comprehensive speech quality analysis. Several open-source toolkits, such as VERSA, AudioLDM-Eval, and FADTK, adopt this strategy to ensure detailed quality profiling~\cite{watanabe18_interspeech, zhang2023amphion, liu2023audioldm, gui2024adapting, shi2024versa}. However, these toolkits often suffer from inefficiencies due to the complexities of using separate models/algorithms and various constraints from algorithmic setups or constraints.

To this end, we introduce \textit{Uni-VERSA}, a unified network designed to assess speech signals comprehensively. \textit{Uni-VERSA} aims to evaluate speech from multiple perspectives by incorporating diverse objective metrics (e.g., signal-to-noise ratio, mean opinion scores, word error rate by a speech recognition model, etc.) within a unified modeling paradigm. Our contributions are as follows: (1) we formalize the \textit{Uni-VERSA} framework, detail its evaluation protocol, and discuss its potential applications across a range of scenarios, including the evaluation of \underline{speech enhancement} systems, \underline{speech synthesis} evaluation, and \underline{conversational speech quality control}. (2) to demonstrate its efficacy, we introduce a benchmark derived from the URGENT24 speech enhancement challenge~\cite{zhang24h_interspeech} and establish a strong baseline by leveraging self-supervised representations. (3) with extensive experiments, our results indicate that \textit{Uni-VERSA} not only provides a comprehensive and efficient evaluation of speech signals but also outperforms methods focusing on a single aspect of human-perceived speech quality.\footnote{A list of pre-trained models are available at \scriptsize{\url{https://huggingface.co/collections/espnet/universa-6834e7c0a28225bffb6e2526}}}

\section{Uni-VERSA}
\label{sec:uni-versa}



\begin{table}[]
    \centering
    \caption{A summarization of metrics in Uni-VERSA. The reference column indicates whether the metric needs reference signals to compute. Detailed reasoning for the selection is discussed in Sec.~\ref{ssec:data-prep}.}
    \vspace{-10pt}
        \resizebox {\linewidth} {!}{
    \begin{tabular}{l|c|c|c}
    \toprule
    Domain & Metric & Range & Reference Type \\
    \midrule
        \multirow{3}{*}{Noise level} & SI-SNR & [-inf, inf] & Signal  \\
         & PESQ & [\hphantom{-i }1, 4.5] & Signal \\
         & DNSMOS & [\hphantom{-i }1, \hphantom{4.}5] & / \\
         \midrule
         Prosody & F0-CORR & [\hphantom{i }-1, \hphantom{4.}1] & Signal \\
         \midrule 
         \multirow{3}{*}{Naturalness} & MOS & [\hphantom{-i }1, \hphantom{4.}5] & /\\
         & UTMOS & [\hphantom{-i }1, \hphantom{4.}5] & / \\
         & SHEET-base & [\hphantom{-i }1, \hphantom{4.}5] & / \\
         \midrule
         \multirow{3}{*}{Intelligibility} & WER & [\hphantom{-i }0, inf] & Text\\
         & STOI & [\hphantom{-i }0, \hphantom{4.}1] & Signal \\
         & SBERT & [\hphantom{-i }0, \hphantom{4.}1] & Signal \\
         \midrule
         Speaker characteristics & SPK-SIM & [\hphantom{i }-1, \hphantom{4.}1] & Signal  \\
    \bottomrule
    \end{tabular}
}
\vspace{-15pt}
    \label{tab:metrics}
\end{table}

\subsection{Metrics in Uni-VERSA Setup}
\label{ssec:data-prep}

Previous studies in speech profiling have often focused on a single target domain, such as noise level, naturalness, or emotion~\cite{cooper2024review, kumar2023torchaudio}, largely due to the scarcity of comprehensive metadata and labels. To overcome this limitation and achieve universality in speech profiling, \textit{Uni-VERSA} leverages a diverse set of metrics derived from both expert models and ground truth annotations, ensuring a robust data preparation process, as indicated in Table~\ref{tab:metrics}. While our ultimate goal is to support all aspects of speech, we begin by focusing Uni-VERSA on five key categories: noise level, prosody information, naturalness, intelligibility, and speaker characteristics. Our initial \textit{Uni-VERSA} benchmark is set up over 11 metrics spanning the five domains.

\noindent \textbf{Noise Level}: This domain assesses speech quality with respect to noise and distortions using three metrics, including \textit{scale-invariant signal-to-noise ratio (SI-SNR)}\cite{le2019sdr, luo2018tasnet}, \textit{perceptual evaluation of speech quality (PESQ)}\cite{rix2001perceptual}, and \textit{deep noise suppression mean opinion score (DNSMOS)}~\cite{reddy2021dnsmos}. Together, these metrics provide a comprehensive assessment of noise levels, with DNSMOS offering perceptual alignment to human judgments, PESQ focusing on reference-based quality comparisons, and SI-SNR quantifying signal-level improvements.\footnote{The use of multiple metrics within a single domain (e.g., noise-level) aligns with Torchaudio-Squim~\cite{kumar2023torchaudio}, which demonstrates that multi-task learning with multiple metrics can improve individual metric predictions. Here, we build Uni-VERSA on the multi-metric prediction idea of Torchaudio-Squim but expand it beyond a single domain.}

\noindent \textbf{Prosody Information}: Prosody, the rhythm, and intonation of speech, differentiates natural speech from monotonous, robotic outputs. We employ the F0 Pearson correlation coefficient (F0-CORR) between enhanced and reference speech as our primary prosody metric. While F0-CORR does not capture all prosodic elements (e.g., duration or stress patterns), it remains the most interpretable measure of pitch dynamics.

\noindent \textbf{Naturalness}: This category evaluates how human-like the speech output sounds using three complementary metrics, including human-annotated MOS, UTMOS~\cite{saeki22c_interspeech}, and SHEET-base~\cite{huang2024mos}. By integrating these metrics, we balance subjectivity, scalability, and generalization in evaluating speech naturalness.


\noindent \textbf{Intelligibility}: To assess how well speech can be understood, we incorporate three metrics, including \textit{word error rate (WER)} by an automatic speech recognition (ASR) system, \textit{short-time objective intelligibility (STOI)}~\cite{taal2010short}, and \textit{Speech BERT Score~(SBERT)}~\cite{saeki24_interspeech}. This multi-faceted approach combines ASR-based, acoustic, and representation-based evaluations for a comprehensive assessment of intelligibility.


\noindent \textbf{Speaker Characteristics}: This domain measures how closely processed speech retains the original speaker's characteristics. We use speaker embedding cosine similarity~\cite{bai2021speaker}, which compares embeddings to determine how well the unique vocal traits of the speaker are preserved.

\subsection{Formulation}
We denote the $i^{\text{th}}$ paired sample $(\mathbf{S}^i, Y^i) \in \mathcal{D}$ in the Uni-VERSA database $\mathcal{D}$, where $\mathbf{S}^i$ is a single-channel speech signal sequence, $Y^i$ is the set of quality metrics, and $i$ is the utterance index. Let $Y^i=\{y^i_b\}_{b \in \mathcal{B}}$ where $y^i_b$ stands for the metric $b \in \mathcal{B}$. Here, $\mathcal{B}$ represents the set of speech profiling dimensions. Some metrics require reference information, which may be involved as auxiliary conditions such as reference audio $\mathbf{S}^{i}_{\text{ref}}$ and reference transcription $\mathbf{T}^i$,\footnote{While in this work, the reference audio is only studied with matching reference, it can be non-matching reference or enrollment speech when it is used in non-matching metrics~\cite{ragano2024nomad, ragano2024scoreq} or speaker similarity.} as listed in the ``Reference Type" column of Table~\ref{tab:metrics}. The main process of Uni-VERSA is defined as:
\begin{equation}
\label{eq:main}
    \hat{Y}^i = \mathrm{UniVERSA}(\mathbf{S}^i, (\mathbf{S}^{i}_{\text{ref}}, \mathbf{T}^i)),
\end{equation}
where $\mathrm{UniVERSA}(\cdot)$ is the Uni-VERSA model and $\hat{Y}^i$ is the set of predicted metrics. Notably, reference signals $\mathbf{S}^i_{\text{ref}}$ and $\mathbf{T}^i$ are optionally used, depending on the design of the model and the availability of the dataset.

The optimization target of Uni-VERSA is rather simple, defined as the $n$-norm of the prediction error between $\hat{Y}^i$ and $Y^i$:
\begin{equation}
    L_\mathcal{B}^i = \sum_{b \in \mathcal{B}}|| y_b^i - \hat{y}_b^i ||_n,
\end{equation}
where in our exploration, we set $n=1$.

The design of the Uni-VERSA introduces several core challenges, stemming from the need to integrate multiple evaluation domains, handle metrics with diverse scales, and contend with limited data availability: \textbf{(1) Integrating Diverse Domains}. As shown in Table~\ref{tab:metrics}, Uni-VERSA spans a wide range of evaluation metrics, from objective, noise-level measures like SI-SNR to subjective naturalness assessments like MOS. A robust model must bridge these varied domains, effectively capturing both the physical attributes of sound and its higher-level perceptual qualities, including linguistic content. \textbf{(2) Handling Varying Metric Scales and Distributions}. As illustrated in Table~\ref{tab:metrics}, the metrics used in Uni-VERSA differ significantly in their numerical distributions. This heterogeneity complicates the task of achieving reliable predictions across all metrics. \textbf{(3) Semi-Supervised Learning Challenges}. In real-world applications, obtaining labeled ground-truth data for metrics like MOS, speaker similarity, or intelligibility is often both scarce and expensive. This scarcity necessitates robust semi-supervised techniques. Additionally, several metrics depend on paired data as indicated in Table~\ref{tab:metrics}. For example, PESQ requires reference audio and WER requires reference transcriptions. In many scenarios, such references are unavailable, forcing models to estimate quality without direct comparisons, which increases prediction uncertainty.

\subsection{Performance Evaluation of Uni-VERSA}

We adopt the evaluation metrics established in previous work~\cite{cooper2024review}, considering both average error and correlation coefficients between ground truth metrics $\{y_b^1, ..., y_b^{|\mathcal{D}|}\}$ and the prediction targets $\{\hat{y}_b^1, ..., \hat{y}_b^{|\mathcal{D}|}\}$. Our evaluation suite comprises linear correlation coefficient (LCC) and Spearman rank correlation coefficient (SRCC).\footnote{The Kendall Tau rank correlation coefficient (KTAU) is commonly used in the literature~\cite{huang22f_interspeech, cooper2023voicemos, huang2024voicemos} and can serve as an evaluation metric. However, in our experiments, KTAU results generally align with SRCC and do not affect the conclusions. Therefore, we simplify our discussion by excluding this metric.} Each metric addresses distinct real-world needs: ranking-based metrics (SRCC) are particularly useful for TTS and speech enhancement evaluations by mitigating range bias (see Section~\ref{sec:intro}), while the linear metric (i.e., LCC) is more suitable for standardizing audio quality control. We also emphasize utterance-level evaluation to ensure the broad applicability of our benchmark across various use cases.

\begin{figure}
    \centering
    \includegraphics[width=0.8\linewidth]{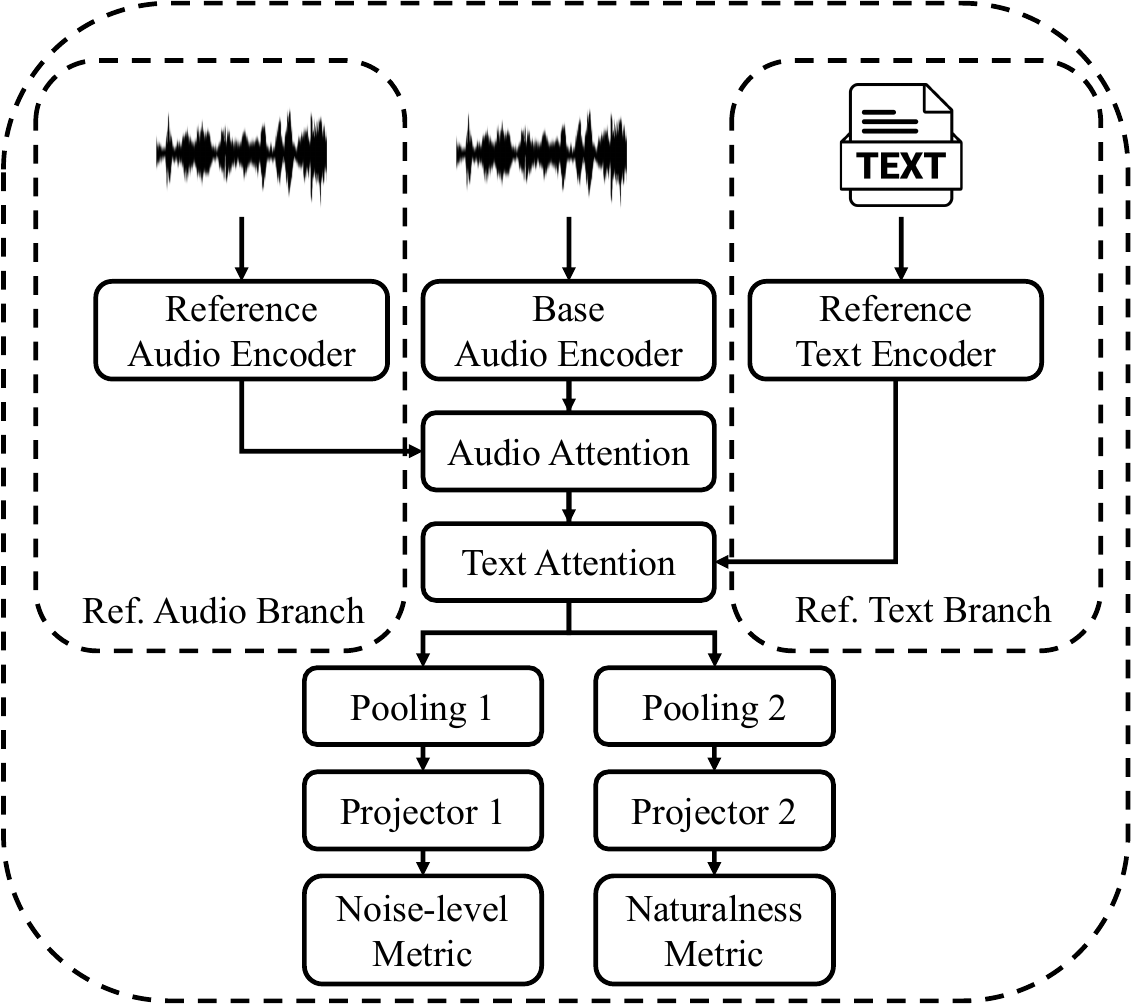} 
    \caption{The architecture of the Uni-VERSA base model with two metrics as an example. Detailed implementation is discussed in Sec.~\ref{ssec: base model}.}
    \label{fig:uni-versa-model}
        \vspace{-20pt}
\end{figure}

\subsection{Baseline Model}
\label{ssec: base model}

\noindent \textbf{Base Model Architecture}. Following the formulation in Eq.~\eqref{eq:main}, we design the base model architecture (i.e., $\mathrm{UniVERSA(\cdot)}$) following the self-supervised learning (SSL)-based speech quality predictor~\cite{cooper2022generalization}. The full-reference architecture, shown in Fig.~\ref{fig:uni-versa-model}, comprises a feature extractor, three distinct encoders, two cross-attention modules, and multiple predictors, one for each metric. The three encoders process the target audio $\mathbf{S}^i$ for quality estimation, the reference audio $\mathbf{S}^i_{\text{ref}}$, and the reference text $\mathbf{T}^i$, respectively. Two cross-attention modules then align the hidden states of the reference audio and text with the target speech representations. Finally, the resulting hidden states are fed into a series of predictors, each estimating a specific metric $y_b$. Each predictor consists of a simple mean pooling layer followed by a linear projection.

\noindent \textbf{Training Strategies}: In real-world scenarios, reference signals $\mathbf{S}^i_{\text{ref}}$ and $\mathbf{T}^i$ are not always available. Consequently, we cannot collect dependent metrics requiring reference signals—such as SI-SNR, PESQ, F0-CORR, STOI, and SBERT for the data. Similarly, subjective MOS labels, due to the expensive human labor, could be only available for a subset of the dataset. Therefore, the training of Uni-VERSA models naturally needs to adopt a semi-supervised strategy. 

Practically, for missing reference signals, we use a 1-second zero-padded audio clip as a placeholder for audio reference and a pseudo transcription from an ASR model for transcription reference. This strategy is applied to both training and inference. During training, if a related metric is unavailable for a given utterance, we simply mask the corresponding predictor.

\section{Experiments}

\subsection{URGENT24 Benchmark}

The Universality, Robustness, and Generalizability for EnhancemeNT (URGENT) 2024 challenge, hosted at NeurIPS2024, is a speech enhancement competition. We construct the Uni-VERSA benchmark by curating a new dataset from submissions to the URGENT24 challenge~\cite{zhang24h_interspeech}.  To mitigate over-tuning contamination, we exclusively use the blind test set submissions. This set comprises 1,000 speech samples: half are simulated (with corresponding reference clean speech) and half are drawn from real-world recordings (without reference signals). This balanced dataset enables thorough evaluation by offering both paired and non-paired speech. Besides, it also features various samples from a collection of state-of-the-art speech enhancement models. Furthermore, the availability of real mean opinion scores ensures that the evaluation closely reflects human perceptual judgments. Accounting for resubmissions, the final collection includes the noisy source speech, the enhanced output from the official baseline, and 112 additional system submissions, totaling 114,000 samples (approximately 293 hours of speech).

Next, we annotate the curated dataset using VERSA~\cite{shi2024versa} with the metrics described in Section~\ref{sec:uni-versa}, adhering to their default configuration in VERSA. We leverage a semi-supervised approach as introduced in Sec.~\ref{ssec: base model}. Since half of the noisy data lacks reference audio, the intrusive metrics are only partially labeled. We also incorporate the subjective MOS evaluation from the URGENT24 challenge, which assesses 300 utterances for each of the 23 final-round submissions, half from simulated noisy speech enhancement and half from real-world scenarios.

Finally, we randomly partition the source speech samples into training, development, and test sets in an 85:5:10 ratio. 
All data are in a 16k sampling rate.

\begin{table*}[]
    \centering
    \caption{Baseline models in URGENT24 benchmark with full references. The results are presented in (LCC / SRCC). Results from the best models are in boldface, considering more decimals. }
        \vspace{-10pt}
    \resizebox {\linewidth} {!}{
    \begin{tabular}{l|c|c|c|c|c|c|c|c|c|c|c|c}
    \toprule
        \multirow{2}{*}{Feature} & \multicolumn{3}{|c|}{Noise-Level}  & Prosody & \multicolumn{3}{|c|}{Naturalness} & \multicolumn{3}{|c|}{Intelligibility} & Speaker & \multirow{2}{*}{Avg.} \\
        \cmidrule{2-12}
         & SI-SNR & PESQ & DNSMOS & F0-CORR & MOS & UTMOS & SHEET-base & WER & STOI & SBERT & SPK-SIM & \\
        \midrule
        Fbank & .69 / .73 & .78 / .80  & .90 / .89  & .56 / .51  & .72 / .72 & .92 / .91 & .85 / .85 & .57 / .53 & .79 / .74 & .80 / .73 & .75 / \textbf{.76}  & .76 / .73  \\
        HuBERT & .84 / .83  & .80 / .81  & .90 / \textbf{.89}  & .61 / .56  & .78 / .77  & .97 / .95  & .93 / .93  & .74 / .71  & .92 / .89  & .94 / .89  & .76 / .71  & .83 / .80  \\
        MR-HuBERT & .85 / .85  & \textbf{.83} / \textbf{.85}  & .90 / .88  & \textbf{.65} / .60  
        & \textbf{.78} / \textbf{.78} & \textbf{.97} / \textbf{.95}   & \textbf{.94} / \textbf{.94} & .74 / .74  & .91 / .89  &  .94 / .90  & \textbf{.79} / .75  & .84 / .82  \\
        WavLM & \textbf{.85} / \textbf{.85}  & .82 / .84  & \textbf{.90} / .88  & .64 / \textbf{.61}  & .77 / .77  & .97 / .94  & .93 / .94  & \textbf{.79} / \textbf{.79}  & \textbf{.93} / \textbf{.89}  & \textbf{.95} / \textbf{.91}  & .79 / .74  & \textbf{.84} / \textbf{.82} \\
         \bottomrule
    \end{tabular}
    }

    \label{tab:full-reference-result}
\end{table*}

\begin{table*}[]
    \centering
    \caption{Baseline models in URGENT24 benchmark with partial references. The results are presented in (LCC / SRCC). The base model is the WavLM-based Uni-VERSA baseline. Results from the best models are in boldface, considering more decimals.}
        \vspace{-10pt}
    \resizebox {\linewidth} {!}{
    \begin{tabular}{l|c|c|c|c|c|c|c|c|c|c|c|c}
    \toprule
        \multirow{2}{*}{Model} & \multicolumn{3}{|c|}{Noise-Level}  & Prosody & \multicolumn{3}{|c|}{Naturalness} & \multicolumn{3}{|c|}{Intelligibility} & Speaker & \multirow{2}{*}{Avg.} \\
        \cmidrule{2-12}
         & SI-SNR & PESQ & DNSMOS & F0-CORR & MOS & UTMOS & SHEET-base & WER & STOI & SBERT & SPK-SIM & \\
        \midrule
        No Ref.  & .84 / .84  & \textbf{.85} / \textbf{.86}  & .90 / .89  & .64 / .60  & .77 / .78  & .97 / .94 & \textbf{.94} / \textbf{.94}  & .79 / .78  & .90 / .86  & .93 / .90  & .79 / .75  & .85 / .82 \\
        Audio Ref. & .84 / .82  & .82 / .83  & .90 / .89  & \textbf{.66} / \textbf{.62}  & .78 / .78  & \textbf{.97} / \textbf{.95}  & .94 / .94  & .78 / .76  & .92 / .89  & .95 / .92  & \textbf{.81} / .77  & .85 / .82  \\
        Text Ref.  & .84 / .83  & .84 / .86  & \textbf{.90} / \textbf{.89}  & .64 / .57  & \textbf{.79} / \textbf{.79}  & .97 / .94  & .93 / .93  &  .78 / .76  & .92 / \textbf{.90}  & \textbf{.96} / \textbf{.93}  & .80 / \textbf{.78}  & \textbf{.85} / \textbf{.83}  \\
        Full Ref. & \textbf{.85} / \textbf{.85}  & .82 / .84  & .90 / .88  & .64 / .61  & .77 / .77  & .97 / .94  & .93 / .94  & \textbf{.79} / \textbf{.79}  & \textbf{.93} / .89  & .95 / .91  & .79 / .74  & .84 / .82  \\
         \bottomrule
    \end{tabular}
    }
            \vspace{-15pt}
    \label{tab:partial-reference-result}
\end{table*}

\begin{table}[t]
    \centering
    \caption{Performance comparison between single-task learning and multi-task learning on MOS prediction.}
        \vspace{-10pt}
    \resizebox {0.6\linewidth} {!}{
    \begin{tabular}{l|c|c|c}
    \toprule
        Feature & Pred. Type & LCC & SRCC  \\
        \midrule
        \multirow{2}{*}{Fbank} & Single & .53 & .55  \\
        & Joint & .72 & .72  \\
        \midrule
        \multirow{2}{*}{WavLM} & Single & \textbf{.80} & \textbf{.80} \\
        & Joint & .77 & .77  \\
         \bottomrule
    \end{tabular}
    }
    \label{tab:joint}
\end{table}

\begin{table}[]
    \centering
    \caption{Out-of-domain evaluation in TTS and conversational data. The base model is the WavLM-based Uni-VERSA baseline without reference signals. The result in the enhancement domain is from the URGENT24 benchmark.}
        \vspace{-10pt}
    \resizebox {0.9\linewidth} {!}{
    \begin{tabular}{l|c|c|c|c}
    \toprule
        Domain & DNSMOS & UTMOS & SHEET & MOS \\
        \midrule
        Enhancement & .90 / .89 & .97 / .94 & .94 / .94 & .77 / .78  \\
        \midrule
        TTS  & .53 / .49   & .65 / .65  & .74 / .74 & .61 / .60   \\
        Conversation & .45 / .45  & .69 / .71  & .60 / .60  & - / -   \\
         \bottomrule
    \end{tabular}
    }
                \vspace{-15pt}
    \label{tab:tts-result}
\end{table}

\subsection{Experimental Setup}

\noindent \textbf{Implementation and Training Details}. We use ESPnet as the training framework~\cite{watanabe18_interspeech} and adopt the frozen WavLM-large model as our feature extractor~\cite{chen2022wavlm}, applying a layer-wise weighted summation as the input feature via S3PRL~\cite{yang21c_interspeech}. Each of the three encoders employs a four-layer transformer architecture with four-head self-attention (256-dimensional), linear layers with 1024 dimensions, and a dropout rate of .1. The text is modeled using 500 vocabulary byte-pair encoding (BPE) units. By default, metric prediction is performed using mean pooling followed by a linear projection.

All models are trained for 50 epochs with a batch size of 16. We use the AdamW optimizer with a learning rate of 0.001 and a linear warm-up scheduler with 25,000 warm-up steps.

\noindent \textbf{Ablation Studies}. In addition to baseline training, we conduct several ablation experiments to evaluate the impact of each module and the usefulness of different reference signals (i.e., audio and text). For the feature extractor, we compare mel filter bank spectral features\footnote{We follow the default Fbank setup in ESPnet~\cite{watanabe18_interspeech}.}, HuBERT~\cite{hsu2021hubert}, and MR-HuBERT~\cite{shi2024multiresolution}.\footnote{While WavLM emphasizes robustness in speech signal modeling, we include MR-HuBERT to assess the potential benefits of complementary multi-resolution SSL representations for the task.} All SSL-based features remain frozen and are incorporated as input features through weighted layer-wise summation. In addition, we conduct additional ablation studies on Uni-VERSA models by completely removing ground truth transcription, reference speech, or both simultaneously.\footnote{Different from the semi-supervised strategy, here we completely remove the reference signals in training.} For systems without reference inputs, we eliminate the corresponding encoder and cross-attention module. To evaluate the impact of joint prediction across multiple metrics, we compare the performance of Uni-VERSA models with models that predict only MOS scores.

\noindent \textbf{Additional Analysis}. Beyond ablation studies, we perform additional analyses to explore the Uni-VERSA system’s applicability in real-world scenarios. Specifically, since the URGENT challenges focus on speech enhancement, we further examine the pre-trained models in out-of-domain settings, such as speech synthesis and conversational speech quality control. We evaluate Uni-VERSA’s performance by comparing its predictions to the original metrics in these contexts and also compare the efficiency of applying Uni-VERSA systems for the problem.

\subsection{Results and Analysis}


\noindent \textbf{Effect of Input Features}. Table~\ref{tab:full-reference-result} shows that SSL representations generally boost prediction performance. While MR-HuBERT and WavLM yield mixed results, WavLM achieves slightly better average LCC and SRCC. UTMOS is the easiest metric to predict (0.95 SRCC with MR-HuBERT-based model), whereas F0-CORR is the most challenging (up to 0.61 SRCC with WavLM-based model). 
This remarks the necessity of balancing the prediction across metrics for future improvement.

\noindent \textbf{The Effect of Partial Reference Audio/Transcription}. The ablation experiments on partial reference are shown in Table~\ref{tab:partial-reference-result}. Interestingly, with sufficient training and domain consistency, models without certain references sometimes outperform those with full references. For instance, the model using reference text shows lower WER prediction correlations, while the model without audio references yields the best PESQ prediction, 
despite audio references theoretically aiding PESQ recovery.
This may stem from the Uni-VERSA model overfitting to training patterns when given redundant reference signals. A potential future direction is to scale up the training data and enhance model generalization (e.g., through regularization or domain adaptation techniques) so that the model avoids overfitting and maintains balanced performance across different metrics.

\noindent \textbf{The Effect of Joint-Prediction}. The ablation experiments on joint prediction are presented in Table~\ref{tab:joint}. Here, Single and Joint refer to the model trained to predict only MOS scores and the model trained to predict diverse metrics simultaneously. While joint prediction improves MOS estimation in Fbank-based systems, WavLM-based models achieve superior performance when predicting MOS alone. These findings suggest that introducing additional prediction dimensions requires careful consideration of both potential benefits and drawbacks. 

\noindent \textbf{Further Discussion}. To further evaluate system performance, we assess the model in out-of-domain scenarios, including TTS data evaluation and conversational speech quality control.
For TTS evaluation, we use the VoiceMOS2022 challenge test set~\cite{huang22f_interspeech}. For conversational quality control, we rely on a self-collected dataset comprising 700 hours of real conversational speech. The results are presented in Table~\ref{tab:tts-result}.

Overall, the pre-trained Uni-VERSA model maintains consistent quality estimation for metrics from expert models and aligns well with human preferences in TTS evaluation, even across different domains (e.g., enhancement versus TTS). 
As an added benefit, Uni-VERSA significantly improves efficiency over direct metric computation with VERSA, for example, achieving up to a 109× speedup on the URGENT24 test set.

\section{Conclusion}
In this work, we introduced Uni-VERSA for multi-dimensional speech quality analysis, detailing its formulation, evaluation criteria, and neural architecture. Our experiments on the URGENT24 benchmark and out-of-domain scenarios demonstrate its effectiveness for various applications. Future work will extend Uni-VERSA to broader speech profiling and other acoustic domains such as general audio and music.

\section{Acknowledgments}

This work is supported by the Defence Science and Technology Agency (DSTA) in Singapore. We would like to thank Daniel Leong and Megan Choo for their valuable comments. Experiments of this work used the Bridges2 at PSC and Delta/DeltaAI NCSA computing systems through allocation CIS210014 from the Advanced Cyberinfrastructure Coordination Ecosystem: Services \& Support ACCESS program, supported by National Science Foundation grants 2138259, 2138286, 2138307, 2137603, and 2138296.



\bibliographystyle{IEEEtran}
\bibliography{mybib}

\end{document}